\documentclass[]{spie}  %

\usepackage{amsmath,amsfonts,amssymb}
\usepackage{graphicx}
\usepackage[colorlinks=true, allcolors=blue]{hyperref}
\usepackage[utf8]{inputenc} %
\usepackage[T1]{fontenc}    %
\usepackage{pifont}
\usepackage{xcolor}         %
\usepackage{booktabs}
\usepackage{hyperref}
\hypersetup{colorlinks,
    linkcolor=red,citecolor=citecolor,urlcolor=blue}
\usepackage{algorithm}
\usepackage{graphicx}
\usepackage{threeparttable}
\usepackage{multirow}
\usepackage{amsmath}
\usepackage{bm}
\usepackage{bbm}
\usepackage{amsfonts}
\usepackage[]{algpseudocode}
\usepackage{enumitem} %
\usepackage[subrefformat=parens,labelformat=parens]{subfig}
\usepackage{colortbl}
\usepackage{geometry}

\usepackage{wrapfig}
\usepackage[algo2e, ruled, vlined, linesnumbered, noend]{algorithm2e}

\definecolor{citecolor}{RGB}{34,139,34}
\definecolor{mydarkblue}{rgb}{0,0.08,1}
\definecolor{mydarkgreen}{rgb}{0.02,0.6,0.02}
\definecolor{mydarkred}{rgb}{0.8,0.02,0.02}
\definecolor{mydarkorange}{rgb}{0.40,0.2,0.02}
\definecolor{mypurple}{RGB}{111,0,255}
\definecolor{myred}{rgb}{1.0,0.0,0.0}
\definecolor{mygold}{rgb}{0.75,0.6,0.12}
\definecolor{myblue}{rgb}{0,0.2,0.8}
\definecolor{mydarkgray}{rgb}{0.,0.2,0.2}

\definecolor{lightred}{RGB}{255,235,235}
\definecolor{lightgreen}{RGB}{235,255,235}
\definecolor{lightblue}{RGB}{235,235,255}
\definecolor{lightcyan}{RGB}{235,255,255}
\definecolor{lightmagenta}{RGB}{255,235,255}
\definecolor{lightyellow}{RGB}{255,255,235}

\definecolor{qxkcolor}{RGB}{215,235,255}
\definecolor{softmaxcolor}{RGB}{230,235,255}
\definecolor{probxvcolor}{RGB}{255,255,235}

\definecolor{topkcolor}{RGB}{255,235,235}
\definecolor{zecolor}{RGB}{255,255,235}
\definecolor{dynacolor}{RGB}{235,255,255}

\definecolor{reviewcolor}{RGB}{0,0,200}

\algdef{SE}[DOWHILE]{Do}{doWhile}{\algorithmicdo}[1]{\algorithmicwhile\ #1}%

\newcommand{\squishlist}{
 \begin{list}{$\bullet$}
  { \setlength{\itemsep}{0pt}
     \setlength{\parsep}{3pt}
     \setlength{\topsep}{3pt}
     \setlength{\partopsep}{0pt}
     \setlength{\leftmargin}{1.5em}
     \setlength{\labelwidth}{1em}
     \setlength{\labelsep}{0.5em} } }

\newcommand{\squishend}{
  \end{list}  }

\title{Toward Large-Scale Photonics-Empowered AI Systems: From Physical Design Automation to System-Algorithm Co-Exploration}

\author[a]{Ziang Yin}
\author[a]{Hongjian Zhou}
\author[b]{Nicholas Gangi}
\author[b]{Meng Zhang}
\author[a]{Jeff Zhang}
\author[b]{Zhaoran Rena Huang}
\author[a]{Jiaqi Gu}
\affil[a]{School of Electrical, Computer and Energy Engineering, Arizona State University}
\affil[b]{School of Electrical, Computer, and Systems Engineering, Rensselaer Polytechnic Institute}

\authorinfo{Further author information: (Send correspondence to Jiaqi Gu)\\Jiaqi Gu: E-mail: jiaqigu@asu.edu}

\begin{document} 
\maketitle

\begin{abstract}
The continued scaling of artificial intelligence workloads is increasingly constrained by data movement, interconnect bandwidth, and energy efficiency in conventional electronic systems. Integrated photonics offers a promising pathway to address these challenges through high-bandwidth optical interconnects and energy-efficient photonic computing primitives. However, translating device-level photonic advances into large-scale, deployable AI systems remains difficult due to strong coupling between physical implementation, system architecture, and learning algorithms.

In this work, we identify three considerations that are essential for realizing practical photonic AI systems at scale: (1) \textbf{dynamic tensor operation support} for modern models rather than only weight-static kernels, especially for attention/Transformer-style workloads~\cite{NP_zhu2024lightening}; (2) \textbf{systematic management of conversion, control, and data-movement overheads}, where multiplexing and dataflow must amortize electronic costs instead of letting ADC/DAC and I/O dominate~\cite{meng_tempo}; and (3) \textbf{robustness under hardware non-idealities} that become more severe as integration density grows~\cite{yin_dac2025}. 
To study these coupled tradeoffs quantitatively, and to ensure they remain meaningful under real implementation constraints, we build a cross-layer toolchain that supports photonic AI design from early exploration to physical realization.
\textbf{SimPhony}~\cite{yin2025dac} provides implementation-aware modeling and rapid cross-layer evaluation, translating physical costs into system-level metrics so architectural decisions are grounded in realistic assumptions. \textbf{ADEPT}~\cite{NP_DAC2022_Gu} and \textbf{ADEPT-Z}~\cite{adeptzjiaqigu} enable end-to-end circuit and topology exploration, connecting system objectives to feasible photonic fabrics under practical device and circuit constraints.
Finally, \textbf{Apollo}~\cite{apollo} and \textbf{LiDAR}~\cite{LiDAR_ISPD_Zhou, LiDAR2_ARXIV_Zhou} provide scalable photonic physical design automation, turning candidate circuits into manufacturable layouts while accounting for routing, thermal, and crosstalk constraints.
Together, these capabilities make our co-design loop both quantitative and physically grounded, bridging architectural intent and deployable photonic hardware.

\end{abstract}


\section{Introduction}
\label{sec:intro}

The rapid scaling of artificial intelligence (AI) workloads has exposed fundamental limitations in conventional electronic computing systems.~\cite{NP_NATURE2017_Shen, NP_ACSPhotonics2022_Feng, NP_DAC2020_Gu, NP_ASPDAC2020_Gu}
While transistor scaling continues to deliver incremental improvements in logic density, system-level performance, and energy efficiency are increasingly constrained by data movement, memory bandwidth, and interconnect power consumption.
These challenges become especially acute in large-scale AI systems, where communication and I/O frequently dominate both latency and energy budgets.

Photonics has emerged as a promising technology to relieve these bottlenecks, offering high bandwidth density, low propagation loss, and natural support for broadcast and wavelength-division multiplexing.
In parallel with advances in optical interconnects and co-packaged/heterogeneous integration~\cite{Tan2023CPO, Ranno2022Packaging}, recent demonstrations have shown photonic computing primitives that accelerate core AI operators (e.g., tensor/matrix computations) at impressive throughput and parallelism~\cite{Feldmann2021TensorCore,Xu2022TensorFlow, meng_tempo, NP_zhu2024lightening, yin_dac2025, NP_Nature_ahmed, NP_NATURE2017_Shen}.
Despite these device- and chip-level successes, a clear path toward \emph{large-scale photonics-empowered AI systems} remains elusive.

A central challenge is that scaling beyond isolated accelerators requires system-level integration across devices, circuits, architectures, interconnect fabrics, and learning algorithms, under constraints that are qualitatively different from electronics.
Photonic integrated circuit (PIC) / electronic photonic integrated circuit (EPIC) implementations must obey curvilinear geometries, limited routing resources, strict fabrication rules, and strong sensitivity to process variation and thermal effects, all of which directly impact loss, crosstalk, tuning power, and yield~\cite{Bogaerts2018ToolsChallenges,SurveyPIC2020}.
Meanwhile, packaging and electronic-photonic interfacing introduce additional constraints and costs that can dominate deployment feasibility and module economics~\cite{Ranno2022Packaging,Tan2023CPO}.
As a result, manual photonic design does not scale to the complexity demanded by system-class AI, and architecture-only abstractions can be misleading unless they explicitly model physical and packaging realities.

In this work, we argue that realizing large-scale photonics-empowered AI systems requires two tightly coupled capabilities:
\begin{itemize}
\vspace{-8pt}
    \item {\textbf{Photonic and electronic-photonic physical design automation (EPDA)}} to enable scalable, manufacturable implementation of complex PICs/EPICs; and
    \vspace{-5pt}
    \item {\textbf{System-algorithm co-exploration}} that incorporates physical non-idealities, control/calibration limits, and packaging/interface costs into architectural design and learning optimization.
    \vspace{-5pt}
\end{itemize}
Drawing on our recent progress, we connect EPDA with cross-layer hardware/algorithm co-design and illustrate how these techniques together enable scalable photonic AI systems.

\section{Photonics-Empowered AI Systems: A Cross-Layer View}
\label{sec:crosslayer}
As argued in Sec.~\ref{sec:intro}, scaling photonics beyond isolated accelerators requires co-optimization across devices, circuits, architectures, and learning algorithms under realistic physical constraints.

In this section, we ground these requirements through three photonic tensor-core (PTC) designs, Lightening-Transformer~\cite{NP_zhu2024lightening}, TeMPO~\cite{meng_tempo}, and SCATTER~\cite{yin_dac2025}, that each target a different bottleneck regime in cloud/edge deployment and demonstrate how cross-layer co-design translates system constraints into implementable architectures.

\subsection{Lightening-Transformer: Dynamic Tensor Operations for Transformer Inference}
To support modern LLMs, particularly attention-based Transformer architectures, photonic computing cores must move beyond weight-static matrix units and enable \emph{dynamic} tensor operations, while jointly optimizing signal conversion and data movement.
Our prior architecture \textbf{Lightening-Transformer}~\cite{NP_zhu2024lightening} was the first photonic accelerator designed to efficiently execute high-throughput, dynamic optical matrix--matrix multiplications for self-attention.
It replaces weight-static photonic matrix units with a \textit{Dynamically-operated Photonic Tensor Core} (DPTC).
At its heart is the \textbf{Dynamically-operated Dot-product (DDot)} engine, a coherent dot-product unit that enables picosecond-level operand switching and supports full-range (signed) matrix inputs without hardware duplication or multiple inference passes.
Lightening-Transformer further integrates these computing cores with \textit{photonic interconnects} for inter-core data broadcast.
By exploiting WDM for spectral parallelism and optical broadcast for operand sharing, the cross-layer-optimized architecture achieves over a 12$\times$ latency reduction compared to prior photonic accelerators.

\subsection{TeMPO: Amortizing Conversion Overheads for Edge-Efficient Photonic AI}
While Lightening-Transformer targets cloud-scale throughput, edge AI faces a different constraint regime where area and energy budgets are highly restricted, and electronic interfaces can dominate total cost.
To address this setting, we extend the dynamic tensor-core concept to \textbf{TeMPO}~\cite{meng_tempo}, introducing an efficient, time-multiplexed dynamic photonic tensor core that improves utilization and amortizes overheads.
At the device level, TeMPO employs customized, foundry-fabricated \textit{slow-light Mach--Zehnder modulators} (SL-MZMs) that leverage enhanced light--matter interaction to achieve a footprint an order of magnitude smaller than standard PDK elements.
At the circuit level, TeMPO tackles the long-standing ADC power bottleneck via \emph{hierarchical partial product accumulation}.
By aggregating photocurrents and using lightweight capacitive temporal integration in the analog domain, TeMPO reduces the required ADC sampling frequency by a factor of $T$ (the integration time step, e.g., 60 cycles).
This cross-layer co-design achieves 1.2~TOPS/mm$^2$ compute density and 22.3~TOPS/W energy efficiency, enabling real-time edge tasks such as voice keyword spotting and semantic segmentation.

\subsection{SCATTER: Robust and Scalable Photonic Tensor Cores under Physical Non-Idealities}
As photonic tensor cores scale in size and density, non-idealities and control constraints (e.g., loss, drift, thermal crosstalk, and calibration limits) become first-order design factors rather than second-order effects.
Our recent architecture \textbf{SCATTER}~\cite{yin_dac2025} exemplifies an extreme cross-layer co-design spanning device, circuit, layout, architecture, and algorithm, where a multi-step co-optimization pipeline jointly targets power/area minimization and robustness under realistic physical constraints.
\ding{202}~Starting from the bottom of the stack, SCATTER replaces communication-oriented foundry building blocks with \emph{compute-tailored} low-power slow-light modulators, enabling substantial baseline reductions in footprint and energy.
\ding{203}~At the physical and circuit levels, SCATTER explores \emph{circuit/weight-matrix co-sparsity} to enable crosstalk-aware layout that safely densifies the photonic tensor core without sacrificing robustness.
\ding{204}~At the architecture level, SCATTER introduces an on-chip \emph{in-situ light redistribution} (rerouting) and power-gating mechanism that dynamically reallocates optical power to active rows/columns, enabling high-efficiency structured sparse matrix multiplication while improving effective SNR by avoiding over-driving inactive channels.
\ding{205}~Finally, SCATTER addresses dominant electronic overhead by upgrading conventional electrical DACs to a \emph{hybrid electronic--optical segmented DAC}, combining high resolution with low power to preserve accuracy at reduced energy.
Together, this cross-stack strategy turns performance, efficiency, and robustness into co-optimized objectives, yielding \textbf{511$\times$ area reduction and 12.4$\times$ power savings} while largely resolving thermal crosstalk, demonstrating a practical path toward robust, sparse, and scalable photonic AI acceleration.

\section{Electronic-Photonic Design Automation: Key Enabler of Scalable Photonic AI Systems}

Section~\ref{sec:crosslayer} demonstrated that achieving high performance, energy efficiency, and robustness in photonic AI accelerators fundamentally requires \emph{cross-layer co-design}, where device physics, circuits, architectures, and learning algorithms are optimized in concert rather than in isolation.
However, while such cross-layer strategies are essential, they also expose a critical scalability challenge: as photonic AI systems grow in complexity, \emph{manual or ad hoc co-design rapidly becomes untenable}.
A photonic AI system's performance is no longer a property of any single layer, but an emergent outcome of \emph{tightly coupled interactions} across the full stack.
For example, device- and circuit-level non-idealities, such as optical loss, thermal crosstalk, directly shape feasible architectures and inference/training strategies.
Conversely, algorithmic choices, sparsity structures, and dataflow patterns feed back into physical layout, routing congestion, and electronic–photonic interface design.

This growing entanglement across abstraction layers motivates a fundamental shift: \textbf{cross-layer co-design must itself be elevated into an automated design paradigm}.
To reliably translate architectural intent and algorithmic innovation into deployable photonic hardware, future photonic AI systems demand \emph{full-stack, physics-aware, and closed-loop design automation}.
In particular, electronic–photonic integrated circuits (EPICs) require design tools that can simultaneously reason about optical and electronic behaviors, propagate physical constraints upward into system-level models, and feed system-level specifications back down to circuits, layouts, and devices.
Electronic-photonic design automation (EPDA) emerges as a key technological enabler to meet this challenge.
In the following sections, we highlight three representative directions from our work that synergistically move toward full-stack EPDA:
(i) \textbf{system-level modeling grounded in rigorous device/circuit/architecture/algorithm co-simulation},
(ii) \textbf{automated multi-objective exploration of photonic circuit topologies under realistic physical and architectural constraints}, and
(iii) \textbf{physically realizable design closure through automated EPIC place-and-route (P\&R)}.

\subsection{SimPhony: Cross-Layer Modeling from Device Response to System Performance}
\begin{figure}[t]
  \centering
  \includegraphics[width=0.75\linewidth]{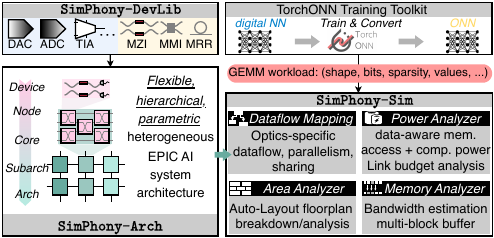}
  \caption{Overview of the SimPhony cross-layer modeling and co-exploration framework. 
  Device- and circuit-level photonic models are integrated with architectural analysis, dataflow mapping, and power, area, and memory estimation. 
  Hardware-aware training and conversion are supported through tight coupling with learning frameworks, enabling system--algorithm co-exploration under realistic physical constraints.}
  \label{fig:simphony_framework}
\end{figure}

Among recent progress in photonic AI system modeling~\cite{Liu2023FIONA, yin2025simphony, cimloop}, we emphasize our open-source cross-layer photonic AI system modeling tool, SimPhony~\cite{yin2025dac}, which serves as the core engine in the evaluation layer in the EPDA stack: it translates device/circuit responses into system-visible metrics so that architectural and algorithmic decisions are made under realistic physical constraints rather than ideal abstractions. 
Figure~\ref{fig:simphony_framework} shows the SimPhony framework, which integrates physical modeling, architectural analysis, and hardware-aware training to support automated system-algorithm co-exploration.
Concretely, this means SimPhony provides a way to propagate device- and circuit-level effects, e.g., loss accumulation, drift, thermal behavior, and calibration/control costs, into the metrics that drive system decisions (throughput, latency, energy, and device non-idealities).

Within the overall workflow, SimPhony acts as the bridge from physical response $\rightarrow$ system response: candidate circuits/topologies are evaluated using device/circuit behavior as inputs, and the resulting system-level cost/performance projections guide which candidates should be further explored and physically implemented. 
This closes the loop between cross-layer intent and the constraints that will ultimately limit deployable hardware behavior.

\subsection{ADEPT: Automated Multi-Objective PIC Topology Exploration}
\label{sec:adept}
With system-level modeling that provides rapid hardware feedback, we leverage it to enable circuit topology exploration to generate high-performance photonic AI hardware.
Existing work has developed automated architecture-level design space exploration~\cite{NP_ICCAD2021_Li} for efficient photonic AI accelerator designs.
As a complementary direction, we have explored photonic tensor core circuit topology optimization using advanced optimization algorithms, ADEPT~\cite{NP_DAC2022_Gu,adeptzjiaqigu}. 
Rather than hand-crafting circuit structures based on heuristics, this framework enables automated Pareto front search over a huge design space, producing Pareto-optimal candidates that can then be embedded in SimPhony as new computing core designs and physically realized by our layout synthesis tools introduced later.

A \underline{key challenge} in photonic tensor core topology exploration is the \emph{exponentially large, highly discrete} design space with multiple competing objectives and constraints (e.g., balancing area, power, latency, robustness, expressivity) that is beyond humans' design capability.

To enable efficient circuit topology design in this massive space \emph{from weeks to hours}, we proposed both differentiable (ADEPT~\cite{NP_DAC2022_Gu}) and multi-objective (ADEPT-Z~\cite{adeptzjiaqigu}) optimization approaches.
The differentiable version relaxes the discrete photonic component selection and construction problem as a soft probability learning problem, leveraging gradient-descent methods for rapid design space exploration while honoring chip footprint constraints.
In this approach, we have successfully found non-intuitive designs that are simultaneously more compact, expressive, and robust compared to prior art, while keeping the whole process within 6 hours.

Building on ADEPT, our latest framework \textbf{ADEPT-Z}~\cite{adeptzjiaqigu} extends to a \emph{gradient-free} optimization formulation that is substantially more flexible than the differentiable relaxation used in ADEPT.
This choice is important for photonic tensor core synthesis because the massive architecture design space is often \emph{highly discrete}, including component types, port configurations, connectivity patterns, and placement decisions, which are difficult to faithfully encode in a differentiable parameterization.
By operating directly in this discrete design space, ADEPT-Z can naturally support richer circuit grammars and constraints, while retaining efficient search.
Equally critical, ADEPT-Z performs \textbf{multi-objective Pareto optimization} that \emph{simultaneously} pushes key system metrics, e.g., \textbf{energy efficiency}, \textbf{compute density}, and \textbf{accuracy/expressivity}, instead of collapsing them into a single scalar objective.
This multi-objective nature empowers ADEPT-Z with the capability of producing \textbf{tens of Pareto-frontier candidates in a single run within $\sim$3 hours}, covering diverse area-power trade-offs that can be carried forward for downstream evaluation and selection.

More broadly, our ADEPT-series illustrates a central insight of EPDA: \textbf{advanced automation and optimization can outperform expert hand-design for complex photonic compute modules}.
By systematically exploring an enormous combinatorial space, it can uncover \emph{non-intuitive} circuit structures that exceed human heuristic design capability, while compressing an exploration process that traditionally takes experts \textbf{weeks} into a \textbf{few-hour} automated workflow.
The resulting Pareto-optimal circuit candidates can then be embedded into system-level models (e.g., SimPhony) and passed to physical implementation tools, enabling a closed-loop path from topology discovery to deployable EPIC designs.

\subsection{Apollo \& LiDAR: Automated EPDA Flow for PICs and EPICs}
\begin{figure}
    \centering
    \includegraphics[width=0.9\columnwidth]{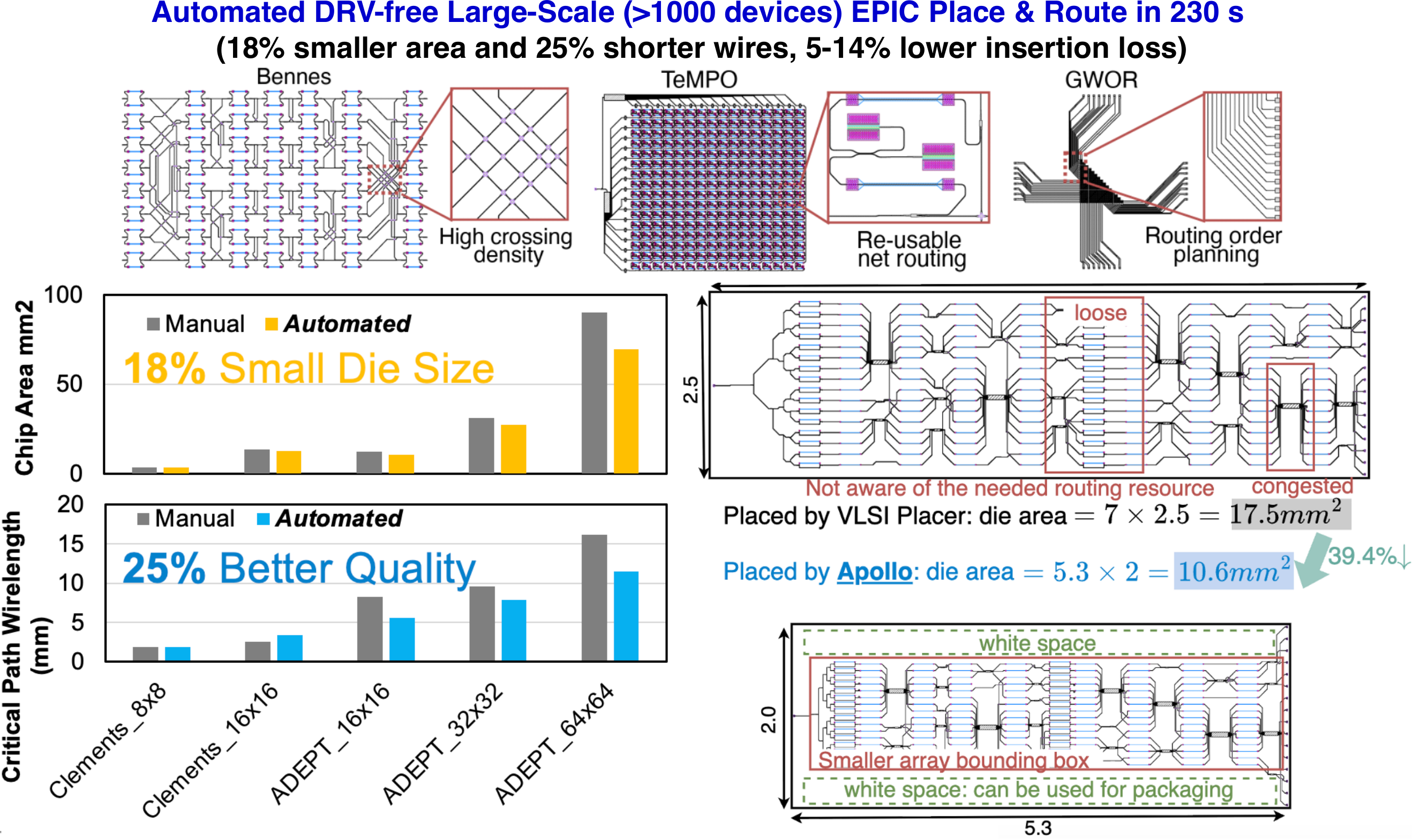}
    \vspace{5pt}
    \caption{Our proposed automated PIC placement engine Apollo~\cite{apollo} and router LiDAR-V2~\cite{LiDAR_ISPD_Zhou,LiDAR2_ARXIV_Zhou} can generate compact, high-quality layout for large-scale PICs (over 1000 devices) within 230s.
    }
    \label{fig:epda_constraints}    
\end{figure}

After topology exploration (Sec.~\ref{sec:adept}) identifies promising circuit netlists, the next bottleneck is design closure: translating these candidates into physically realizable layouts while accounting for curvilinear waveguide constraints, routing congestion, crossings, and layout-induced loss/crosstalk.
Early demonstrations of PIC physical design largely relied on manual, ad hoc methodologies. While sufficient for proof-of-concept devices and small-scale circuits, such approaches do not scale to system-level PICs comprising thousands to millions of components. In practice, PIC development still follows a sequential, largely manual pipeline, device design, schematic capture, layout drafting, and verification~\cite{NP_book_chrostowski}. 
Custom blocks (e.g., modulators, filters, and multiplexers) are typically handcrafted,
assembled at the schematic level, and then translated into layout through manual or rule-based placement and routing. This fragmented workflow significantly slows iteration and becomes a fundamental bottleneck as photonic systems scale.

To overcome this barrier, electronic-photonic design automation (EPDA) is urgently needed for automated PIC layout synthesis.
Recent research has explored automated placement and routing for photonic circuits considering various realistic constraints~\cite{LiDAR_ISPD_Zhou, apollo, Wu2025AutomaticPICRouting,zheng2021topro, zheng2021topro}.
An essential first step is to develop automation that can translate a designer-specified netlist and constraints into a high-quality physical layout by \emph{automatically placing} photonic components while accounting for routability and layout-dependent effects. To this end, we propose \textbf{Apollo}~\cite{apollo}, the first GPU-accelerated, routing-informed placement framework tailored for large-scale PICs.
Rather than placing components independently of routing considerations, we explicitly model waveguide routing congestion and crossings during placement to preserve enough routing spacing for routability maximization.
Figure~\ref{fig:epda_constraints} illustrates how routing-informed placement methodology substantially improves \emph{layout regularity, area efficiency, and routability} in large-scale PICs within only \textbf{230s}.
This approach is essential for large-scale PICs, where naive placement can render routing infeasible.

Following placement, our tool \textbf{LiDAR-V2}~\cite{LiDAR_ISPD_Zhou,LiDAR2_ARXIV_Zhou} executes waveguide routing, distinguishing itself from existing methods by generating design-rule-violation (DRV)-free, GDSII layouts. Unlike traditional approaches that apply post-hoc smoothing, which often fail under congested constraints, LiDAR-V2 integrates a curvy-aware A$^\ast$ search that incorporates node orientation and bending radius directly into neighbor generation. And we propose an orientation-aware bitmap that enforces spacing rules and facilitates dynamic waveguide crossing insertion, eliminating the need for manual planning.
As shown in Fig.~\ref{fig:epda_constraints}, the result remains routable even under high crossing density and successfully generates valid layouts without DRVs while achieving a \textbf{5--14\% insertion loss reduction} compared to the previous work.

In addition, metal routing is often overlooked in research prototypes, yet it can consume a substantial fraction of the overall layout closure time in practice. To this end, we explicitly incorporate metal routing~\cite{zhou2025photonics} into our flow. Compared with conventional VLSI routing, PIC electrical routing must tightly coordinate with photonic structures and waveguides, account for packaging-driven pad breakout and long-distance interconnects, and obey more diverse keep-out and coupling constraints (e.g., to avoid optical loss, crosstalk, and thermal interference). By modeling these PIC-specific requirements, our flow delivers routing solutions that better match designers' needs while remaining scalable and design-rule compliant.

Recent automated frameworks from our group demonstrate that these stages can be unified into a \textbf{push-button EPDA flow}, translating high-level photonic circuit descriptions into fully routed layouts with minimal human intervention. 
Physical design automation is not merely a productivity enhancement; it is a \textbf{key enabler of scalable photonic AI systems}. By automating layout synthesis and enforcing physical feasibility, EPDA fundamentally expands the architectural design space that can be explored within realistic time and resource budgets. As shown in Fig.~\ref{fig:epda_constraints}, iterative placement--routing refinement achieves, on average, \textbf{an 18\% reduction in die size and a 25\% improvement in layout quality}.
Overall, our EPDA toolflow allows system designers to evaluate larger and more complex photonic architectures, explore tighter integration densities, and interconnect strategies while maintaining predictable performance. 

\section{Discussion}

This work suggests that the main scaling bottleneck for photonics-empowered AI is not generating new circuit concepts, but closing the loop from cross-layer intent to implementable hardware under realistic loss, thermal/control, and layout constraints. 
The AI-assisted EPDA flow presented here helps reduce this gap by (i) evaluating system behavior from device/circuit responses, (ii) exploring topologies with implementation constraints in mind, and (iii) using scalable P\&R to enforce physical feasibility.

Two practical directions could further strengthen this loop: (1) packaging/interface-aware modeling integrated earlier in evaluation so coupling and assembly constraints are reflected before topology and layout commitments, and (2) more systematic variability/test hooks (variation-aware metrics and calibration/test planning) so candidate designs are screened for robustness rather than relying on post-fabrication fixes.


\end{document}